\documentclass[a4paper,11pt]{article}
\usepackage{pstricks}
\usepackage{epsfig}
\begin{document}  
  
\title{A Tip of the TOE}  
  
\author{P.~P.~Marvol$^1$}  
\maketitle 
\vspace{-5cm}\hspace{9cm}physics/0309055  
\vspace{4cm} 
 
\begin{center}  
Institute for Applied Quantum Acoustics, \\   
Pyramid College Cacania, Hatetepe Boulevard 12,\\  
CC-12000 Cacania City, CACANIA
\end{center}  
{}\hspace{75mm}{\it In memoriam E. F. Dr\"acker}  
\vspace{0.3cm}  
\begin{center}  
\begin{minipage}{12cm}  
{\it Abstract.} Using  standard methods from string theory,   
this paper presents a comprehensive survey about the most  
important aspects of the theory of sand. Special interest is put on the   
examination of the sand-wind duality and the interaction of ordinary   
(non-supersymmetric) sand with the heat field of the earth, as solution    
of the field inequalities of a stone. This will lead us, in a natural way,  
to completely new insights into the theory of sandstorms.  

\vspace{5mm}
\end{minipage}  
\end{center}  
\footnotetext[1]{marvol@salzamt.at}  
   
\section{Introduction}   
   
The matter of sand has ever since fascinated mankind. Already twenty thousand   
years ago attempts to classify the grainy zoo have been made, \cite{abc}.   
In the Cretacious Age many advances with respect to sand have been   
undertaken,    
especially it has been proved by \cite{unspellable} that chalk and sand    
are---at a fundamental level---two manifestations of the same thing.    
Unfortunately, since the end of the Cretatious Age, sand research activities   
suffered from a considerable lack of financial resources, a time period   
which is called the Big Sand Crisis (BSC). This led to the the   
 well-known revolutionary movement ``Every Scientist Needs   
a Camel'', \cite{marx}. After all, at least a few new concepts have    
been invented, among them the so-called {\it wheel}, which then, however,   
has been discarded due to its impracticability,    
\cite{tuareg}. (However, there are some unteachable fanatics    
who still try to demonstrate the usefulness of wheels on sand in   
the annual Paris-Dakar competition.)   
     
Some centuries after the BSC a completely new impulse   
came from ancient Lei Li Ga, \cite{lei}, who was the first to put   
special interest in the dynamics of sand.   
   
The so far highly disordered efforts in sand have been fibre bundled by   
the famous philosoph Goe-The, who formulated the central   
question of today's complex theory of sand, \cite{fist}:   
   
{\center\it Herauszufinden, was den Sand im Innersten zusammenband.$^{2}$ }  
{}\\   
\footnotetext[2]{{\it to be translated maybe as:}    
To find out, what sand bound together in its inner.}    
   
Starting from this point, many interesting theories about sand dynamics   
have been developed. One of the most ambitious (and ambiguous)   
models was the well-known sandstring theory developed by the old    
Egyptian priest Edua 'Wit X  
\cite{cheops}. With the postulate of the   
non-existence of infinitely small grains of sand, several severe problems   
of ordinary sand-theory (e.g.~incurable sand-divergencies,    
causality problems etc.) seemed to be repaired. However,    
over the millennia, this model did not bear any fruits in sand. 
(After all, there have been some interesting side results for weavers, 
in particular new techniques for knotting.)
This led to a complete restart of research in the   
field of sand. Fortunately, at least some interesting questions have been   
solved independently of sandstring theory, \cite{cameroon}, \cite{unknown},   
\cite{acta}.    
   
Modern, post-string sand theories have also their origin in Egypt. In    
particular, a   
comprehensive quantum theory of the dynamics of sand dunes has been    
developed, \cite{annals}, \cite{quant}.   
Piles of stones (so called ``pyramides'') have been    
constructed for a macroscopic test of the dynamic theory, especially for the   
prominent dune-tunnel effect. However, the    
pyramid-models emerged as too roughly textured, so there could not be    
proven anything with these (although they are still very impressive).     
Indeed, experimental clues for the correctness of the Egyptian Theory of   
Quantum Sand Dynamics (ETQSD) has been found only recently by Swedish    
Inger Zeil, \cite{zei}. Of course, to people living in North-Africa, his results      
have always been obvious. However, as they    
aroused some interest outside the Big Desert, he eventually will receive   
the prize without any bells.    
   
Presently, research in sand dynamics came to an interlocutory end with    
a stone's$^3$ \footnotetext[3]{Unfortunately, the authorship of this work is still  
 unclear.}   
field inequalities,    
\cite{stone}, which form the basis of our approach.    
   
The paper is organized as follows. In section two, the main aspects    
of the standard sand-wind duality are recapitulated. In section three   
a stone's field inequalities (specialized for the heat kernel of   
the earth) are presented. From these fundamental considerations,   
section four will lead us in a natural way to a completely new and    
attractive theory of sandstorms. In section five follows a brief conclusion    
and outlook with respect to some recent aspects of sand theory will be given.   
   
\section{Sand-Wind Duality}   
   
This section is devoted to a brief recapitulation of the sand-wind duality   
of grain dynamics. Since sand is quantized ('grained'), one has to use   
the well-known cat-equation, which reads (in natural units)   
\begin{equation}   
 (W + V_H)|\mathrm{sand}\rangle = -i\frac\partial{\partial t}|\mathrm{sand}   
\rangle.   
\end{equation}   
$W$ is the wind operator, $V_H$ stands for the heat field operator.    
We use the idempotency of sand dunes (two sand dunes thrown upon each other   
give again a sand dune, since the superfluous grains merely drain in the sea   
of sand, as introduced by the great wizard Pamdirac) and $WV_H=V_HW=0$    
(heat and wind never occur at the same time). Now,   
multiplying this equation by its complex conjugate and   
dressing the cat with its bra leads immediately to   
\begin{equation}   
\langle \mathrm{sand}|W|\mathrm{sand}\rangle =    
\langle \mathrm{sand}|(\frac\partial{\partial t})^2\mathrm{|sand}\rangle.   
\end{equation}   
Due to the quadratic time derivative on the right hand side of this    
equation, by interpreting the operators as states and the states as operators   
we can immediately read off the duality between sand and wind. Most    
interesting consequences of this phenomenon follow from the fact that   
a (sand theoretical) distinction between the sand-field and the   
wind-field is impossible. This is the basis of our modern understanding of   
sandstorms.   
   
In classical sand theory, it has been believed that the genesis of sandstorms   
lies merely in the wind field, interacting with the grains of sand, pushing   
them up and around. Modern understanding of sandstorms using the sand-wind   
duality of grain dynamics, however, has taught that the wind is a mere   
manifestation of the sand field dynamics. Thus, the usual theory of    
sandstorms is simply that the interaction of the sand-wind-field    
with itself is sufficient to create a sandstorm. Indeed, it has been shown   
by \cite{private} that the movement of the grains of sand functions as   
source term for the wind field, not the other way as suggested by classical    
theory: It's the grains generating the wind.   
   
So far, this is somwhat awkward but, after all, well understood.    
Unfortunately, combinig the principles of grain dynamics with   
the Relative Generality Theory of a stone, we will find that this simple,   
non-general model is not sufficient to a full understanding of sandstorms.   
     
\section{The Heatfield}   
   
In the preceding section we have deliberately assumed that the reader is familiar   
with some of the basic notions describing the coupling of wind to sand.   
However, in order to fully exploit the origin of the breakdown of sand-wind-duality in   
sandstorms, we shall need a more thorough understanding of the generalistic physics   
of sand with respect to the heat field, i.e.~of a stone's theory and its grainization.   
An excellent recent introduction can be found in    
 \cite{fein} -- in chapter 42, of course.   
\vspace{5mm}  
 
Let's begin with a brief historical survey.  
As it is well known, the experiments of Michael's son (an ethiopian sprinter)  
and a cheetah \cite{mic} have given a first evidence that nothing moves faster   
than sand and that all grains (in a sandstorm) have  the same speed,   
at least within the experimental accuracy of that time. This led to a relatively   
special view of  matters. However, it still remained unclear how such a theory  
could be combined with the interaction of wind, sand and heat in the spirit of the old   
ton theory \cite{cern}.  
Only a few millennia later, this problem has been solved in an ingeniously simple   
manner by a stone.  
  
Obviously, the sand in the desert is rather hot,   
in fact even hotter than the air surrounding it. Up to a stone's  
era many scientists desperately tried to understand this phenomenon.  
A stone turned it into a new paradigm instead. In brief, his idea can be stated as   
follows: The  heat moves the grains, but the grains are the source for the heatfield at   
the same time. (If we could switch off the dynamics and  wait long enough then the air  
would be as hot as the sand.) More quantitatively, a stone's inequalities read  
\begin{eqnarray}  
        T_g^{\mu\nu} &\geq& T_w^{\mu\nu} \quad ,   \\  
        G^{\mu\nu} &\approx& T^{\mu\nu}_{\mbox{?}}  
\end{eqnarray}  
from which one immediately derives his famous saying ``E=m ceh ceh''.          
(The ancient letter $=$ has no analogue nowadays.)  
Note that the sand is still treated as a classical (in the sense of continuous) field   
(of characteristic zero) here.  
The coupling of the spin-2-heatfield to wind, which is spin-1, i.e.~a vector-field,  
is described by the force   
$ F^\mu = h^{\mu\nu} W_{\nu} $ acting on the wind quanta.  
  
Due to its many nontrivial predictions, the success of a stone's theory was  
overwhelming.  
Among these predictions were the delay of the time shown by a sandglass in an   
external heatfield and, most striking, the deflection of light in heatfields, often   
referred to as ``Fata Morgana''.  
(By the way, this effect is also the origin of the common misbelief that   
dromedars have two hunches.)  
  
Even more so, it turned out that the grainization of this theory was no great deal 
$^4$\footnotetext[4]{since a spin-2 grain, i.e.~the heat quanta,   
appears without saying in the   
grainized theory}.  
A typical effect that can only be understood in the framework of the full   
theory of grain dynamics, as it is a typical grain effect of the sand-wind-interaction,  
is the movement of sandpiles \cite{quant}. We have already given a detailed account  
of this application in the preceding section. We have also shown therein that the   
resulting sand-wind duality leads to our present understanding of the genesis of  
sandstorms.  
  
But we also mentioned, that this model has some drawbacks to which we shall turn  
our attention now.  
  
First of all and honestly speaking, the above described mechanism for the genesis   
of sandstorms is only of limited theoretical value. In fact, no one has ever been   
able to prove that such extreme states of the sand-wind-field like sandstorms  
really do exist. (Actually, it is not even  clear,   
whether this theory describes anything realistic at all.)  
A modern researcher can hardly take a ``problem''of this kind  serious, of course,  
but unfortunately it is not the end of the story:  
  
Evidently, there is only a finite amount  of energy in the heatfield of the earth.  
It came as a big surprise, when it has been discovered that certain kinds of  
sandstorm, so called grain ray bursts (GRB) had energies seemingly exceeding this  
upper bound, if a stone's (grainized) theory was valid in this regime.  
  
For a few hundred years this discovery aroused a lot of confusion and   
many scientists   
even started to doubt the spherical shape of the earth, until Re-Ez \cite{reez} remembered  
the forgotten singularity theorems of \epsfig{file=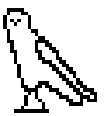,width=0.5cm} 
\epsfig{file=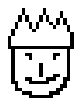,width=0.5cm} and  
\epsfig{file=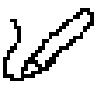,width=0.5cm}$\;$\epsfig{file=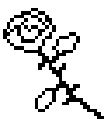,width=0.5cm} 
, \cite{Penhawk}, for the heatfield .  
The former, \epsfig{file=hawk.eps,width=0.5cm} 
\epsfig{file=king.eps,width=0.5cm}, has also shown that such singularities will ``radiate'' large amounts of  
sand.  
  
Assuming that the GRBs stem from the creation of sand due to the presence of   
a singularity of the heatfield he overcame all problems and eventually caused  
the (long overlooked) second revolution in the field of sand.  
  
In fact, it soon became clear that there do exist many visible singularities   
of the heatfield.  
For instance, the sun is nothing but such a singularity, arising periodically  
and then moving at the sky. Due to the emittance of sand$^5$  
\footnotetext[5]{Amusingly,  
this mechanism resulted in the ancient misbelief that the sun is made out of  
yellow sand.}, it disappears again  
after a certain time, depending on the state of the heatfield, i.e.~the temperature.  
(In summer it will last longer than in winter.)  
  
Quite recently, even the existence of a very massive heat-kernel   
in the center of earth has been established.   
  
Most importantly and even more surprisingly, however, it turned out that many ordinary  
(non-GBR) sandstorms can be traced back to such singularities.  
This motivates our conjecture that {\em all} sandstorms are caused by  
such singularities. 
 
\vspace{5mm}   
Finally, the interpretation of the genesis of sandstorms as caused by singularities also   
resolved another puzzle:  
  
Since sand is fermionic (it obviously respects the exclusion principle and, of course,  
the grains have spin $\frac{1}{2}$) and interacts only via heat-interactions,  
the system should be stable, seemingly in contradiction to the existence of sandstorms.  
However, sandstorms are only a local instability in the superfield,   
just like oases \cite{acta}. (For this reason, the energy stored in a sandstorm does not exceed 
the maximal available energy.) 
Globally, the system is in equilibrium!  
This fact might be of some importance.  
  
\section{Sandstorms Revisited}   
So let's assume that in the center of each sandstorm  a singularity of the  
heatfield is sitting.$^6$\footnotetext[6]{The objection that such a singularity  
has never been observed in experiments  
with non-GRBs can be ignored:  
People entering a sandstorm voluntarily should not be taken serious.}  
As it is well known, in the eye of a (sand)storm the wind vanishes,  
and thus it is commonly believed that $WV_H=V_HW=0$ holds  
even there. But that is wrong, since $V_H$ is infinite in the center  
of a sandstorm. Accordingly, the sand-wind-duality breaks down in this region,  
and ``that  is the poodle's core ''\cite{fist}. 
 
Before we can explore this fact and its consequences in detail we should describe  
the mechanism of the emergence of these singularities, however.  
This mechanism is quite similar to the mechanism discovered by Msw the Elder  
\cite{msw}, when he   
tried to understand the so-called solar-new-tiro puzzle.   
(The problem why the sun emits so few sand, which is solved by   
a resonance mechanism. At the same time the Msw-resonance   
explains why the sun emerges only periodically and why it is red around the times of its 
appearance, respectively disappearance.)

Thus, we immediately infer that small fluctuations   
in the heatfield will eventually  lead, by a similar   
resonance, to the emergence of a singularity if we apply the following  
well-established result \cite{myself}: 
 
\vspace{0.5cm}\noindent 
{\bf Theorem 1.}  
Under the above assumptions, there exists a point $x\in \mathbf{R}^3 $ in which the degree   
of singularity of the heatfield-distribution $\langle  
\mathrm{sand} |h^{\mu\nu}   
|\mathrm{sand} \rangle $ is strictly larger than $C_{storm}$.   
 
\vspace{0.5cm}  
Thus, the genesis of a sandstorm will be inevitable. Moreover, it is now evident, that  
the standard sand-wind duality is broken, viz.  
\begin{equation} \lim_{r\to x }\langle \mathrm{sand} \label{result}  
|V_H W |\mathrm{sand} \rangle (r) \neq 0.  
\end{equation}   
Hence, {\bf the grains in a sandstorm  are driven by the wind!}   
This is the most nontrivial and surprising result of this paper. It can and should be verified  
empirically.

\section{Conclusion and Outlook}   
In this paper we presented in a highly suggestive and self-explaining way a new idea  
about the emergence of sandstorms. We have shown that the movement of the grainized   
sand is caused by the wind field. This result seems to be in consistency with the  
predictions of the original (but wrong) classical sand-theory. Note,  
however, that this ``accordance'' is not even qualitative: There are no similarities of  
the classical theory with our completely new and astonishing result (\ref{result}).  
 
Of course, this paper has only sketched some basic features of the new sand theory.  
A complete review is in preparation. There furthergoing questions about  
noncommutative sand, the super-connection between  quicksand and worme holes  
and some quite interesting (although avantgardistic) new features in sand 
theory will be presented \cite{greek}. 
Moreover, we will overcome some of the flaws this paper is suffering from, 
e.g.~a renormalization of the infantilities due to the  
self-citation \cite{myself} will be done. This works the following way:  
Assuming   
that the problem is already removed at the first order. Then it follows from 
an inductive proof that no infantilities will arise at any higher order 
due to recursive citations. Now, curing the problem at the first loop level  
is rather simple. This completes the proof of renormalizability of this work. 
   
We should also admit that this work suffers from a considerable lack of computer   
assistance. Unfortunately, the current generation of the abacus (used here at   
the Pyramid College)  seems unable to perform simulations of such complexity.  
Not to speak of resolving a heart-tree in the fog.

\vspace*{1.5cm}  
\noindent  
{\bf Acknowledgements} \\  
I would like to thank my colleagues Mario Paschke and Volkmar Putz who made me aware  
of the reference \cite{myself}. Without their support and friendship I would not   
exist. Furthermore I have to thank my namesake P.P.~Longstockings, from whom much   
inspiration has been drawn in making this paper:  
{\em ``No one really knows...''}

\vspace{1cm} 
\epsfig{file=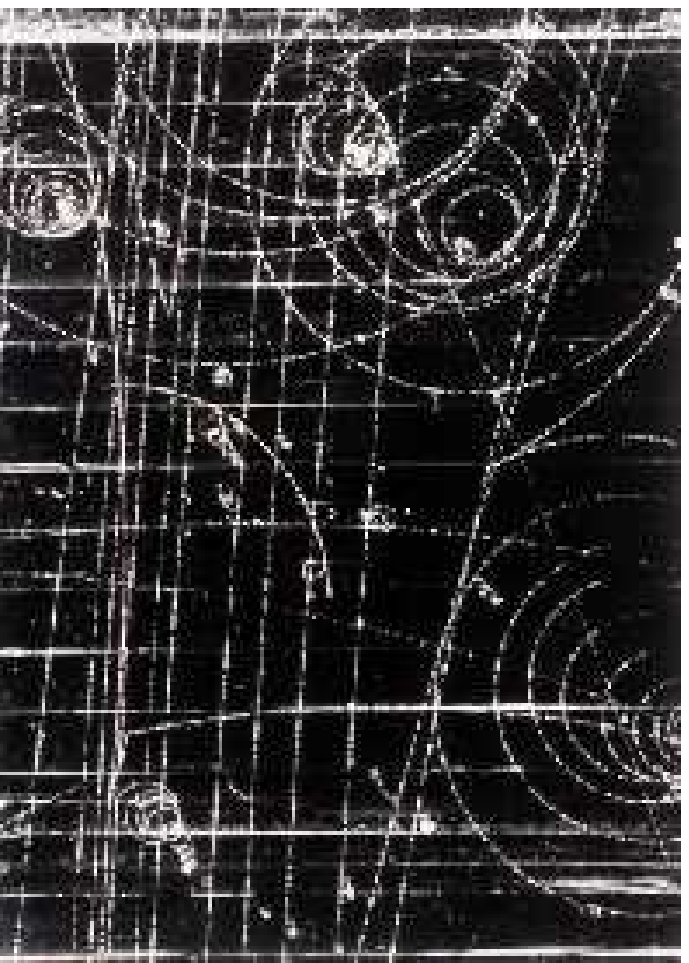,width=9cm} 
 
\vspace{0.5cm} 
Figure 1 (The so-called $\Omega^{- -}$-facsimile). 
 

\begin{thebibliography}{99}   
   
\bibitem{abc}   
see fig.1, facsimile taken from {\bf Sand in our times 15, 199985 \underline{BC} .}  
   
\bibitem{unspellable}$\ \ $ 
 
\vspace{-0.7cm}\epsfig{file=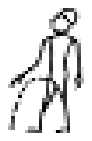,width=1cm}  
\quad. 
 
\bibitem{marx}   
K. Murx, {\em ``Proletariat and Sand,''} work in regress.   
   
\bibitem{tuareg}   
Abou Tsa Id, {\em ``The Wheel and its Possible Applications,'' }   
{\bf Tuareg Journal of Modern Physics A 8000    
\underline{BC},} stone tables 12-12.5.   
   
\bibitem{lei}   
Lei Li Ga, {\em ``But still it moves,''}
{\bf Crackpot Rev. Let. of Common Sand Vol. MCMDXXXIII.}    
   
\bibitem{fist}   
Goe-The, {\em ``Fist I,''} {\bf Reclam excavations,} edited by the police  
department of Leipzig.   
   
\bibitem{cheops}   
Edua 'Wit X, {\em ``M-, F-, Sand-Strings and all that}'',  
{\bf Pyramide of Cheops, corridor 6B (periodically under    
water),} table 1267.  
   
\bibitem{cameroon}    
Shrinkhead No.~7, {\em ``Some least interesting questions of sand''},   
{\bf Came\-roon Jungle Drums 6000 \underline{BC}.}   
   
\bibitem{unknown}   
{\em ``The Grand Unification of Sand and Migraine,''}    
reference unknown, but widely distributed.   
   
\bibitem{acta}   
He, who drank too much of the muddy water (if the hyroglyphes are translated correctly),  
{\em ``Do oases really exist? IV : The unreasonable effectiveness of washing 
 clothes with sand.''},    
{\bf Acta Mesopotamica Vol.  4.8 \%.}    
   
\bibitem{annals}   
Pamdirac, {\em ``Spells on Grain mechanics''}, {\bf Beduin Acad. Press.} 
 
  
\bibitem{quant}  
A most recent review -- though written in a somewhat old-fashioned language   
-- can be found under:  
  
{\bf www.marxist.com/science/dialecticalmaterialism.html}  
   
  
\bibitem{zei}   
I.~Zeil [{\em innumerable references}],  
{\bf Comm. in Priv.  Mat. Vol. 12 - 400.}   
   
   
\bibitem{stone}   
[Author unknown], {\em ``The Relative Generality Theory of the Heat Field,''}   
{\bf found on a large stone near the Sahara.}  
    
\bibitem{private}   
Probably going back to Apu, who had told this to H.~Simpson, who had told it to his wife,   
who had told it to her hair-cutter... 
  
\bibitem{fein}  
{\bf Feynman's Lectures on Physics, Vol.~II, section 42.1.}  
  
\bibitem{mic}  
 {\em ``The ultimate the-looser-looses-the-head race, final results,''}   
{\bf Proc. Emp. Soc. of Sand,  around breakfast.}  
  
\bibitem{cern}  
1237 anonymous authors, {\em ``On Temperature and Evaporation of apples stored in   
new and old tons,''}  {\bf Clear and Evident Research  
of Nubia,} reprint of sheepskins 11-27.   
  
\bibitem{reez} 
Re-Ez, {\em ``I don't get the problem ''} contribution to the  
{\bf 2. Masai-meeting of heat and wind}  2000 \underline{BC}. 
 
\bibitem{Penhawk} 
\epsfig{file=hawk.eps,width=0.5cm} \epsfig{file=king.eps,width=0.5cm} \&  
\epsfig{file=pen.eps,width=0.5cm}$\;$\epsfig{file=rose.eps,width=0.5cm} 
{\em ``The large grain structure of sand''}, {\bf Severe Lectures in Sand,} 
Knight, 12 \underline{AD}.  
 
\bibitem{msw}  
Msw the Elder  {\em ``What the great god Fe y Nman told me,''}  
 {\bf Annals of Nil-Pollution,  Thebia-Series,} pp. 387-402.   
  
\bibitem{myself}  
P.P.Marvol  {\em ``A Tip of the TOE,''} physics/0309055.  
  
\bibitem{greek} 
X.~Inachos \& Y.~Theseus, ``{\em A Brief Introduction to Brain Theory,}''  
{\bf Artemis-Temple-Library of Athens},  
destroyed by the Spanish Inquisition.  
\end{thebibliography}
\end{document}